%% file: metro.tex
\documentclass[12pt]{article}

\usepackage{epsfig}
\usepackage{cite}
\usepackage{inputenc}
\inputencoding{latin1}

\def\f2{F_2}
\def\tf2{$\f2$}
\def\q2{Q^2}
\def\tq2{$\q2$}
\def\qt2{{q_\perp^2}}
\def\tqt2{$\qt2$}

\def\k2t{k_\perp^2}
\def\tk2t{$\k2t$}

\def\l2qcd{{\Lambda^2_{\mbox{{\scriptsize QCD}}}}}
\def\t2lqcd{$\l2qcd$}

\newcommand{\JCP}[3]{{\it J.~Chem.~Phys.} {\bf #1} ({#3}) {#2}}

\newcommand{\CPC}[3]{{\it Comput.~Phys.~Comm.} {\bf #1} ({#3}) {#2}}

\begin{document}

\begin{titlepage}

  \begin{flushright}
    DESY 99-133\\
    RAL-TR-1999-061\\
    TSL/ISV-99-0216\\
    September 1999
  \end{flushright}
  \begin{center}

    \vskip 6mm
    {\LARGE\bf The Metropolis algorithm for on-shell\\[0.35cm]
      four-momentum phase space}
    \vskip 8mm

    {\large Hamid Kharraziha}\\[0.25cm]
    {\it II. Institut f\"ur Theoretische Physik}\\
    {\it Universit\"at Hamburg}\\
    {\it D-22761 Hamburg, Germany}\\
    hamid.kharraziha@desy.de\\
    \vskip 6mm

    {\large Stefano Moretti}\\[0.25cm]
    {\it Rutherford Appleton Laboratory}\\
    {\it Chilton, Didcot, Oxon OX11 0QX, UK}\\
    moretti@v2.rl.ac.uk\\[3mm]
    and\\[3mm]
    {\it Department of Radiation Sciences}\\
    {\it Uppsala University, P.O. Box 535}\\
    {\it 75121 Uppsala, Sweden}\\
    moretti@tsl.uu.se

  \end{center}
  \vskip 6mm
  \begin{abstract}
    \noindent
    We present several implementations of the Metropolis method, 
    an adaptive Monte Carlo algorithm,
    which allow for the calculation of multi-dimensional     
    integrals over arbitrary on-shell four-momentum phase space. 
    The Metropolis technique reveals itself very
    suitable for the treatment of high energy processes in 
    particle physics, particularly when the number of final state objects
    and of kinematic constraints on the latter gets larger. We compare the 
    performances of the Metropolis algorithm with those of other programs  
    widely used in numerical simulations.
  \end{abstract}

PACS: 12.20.Ds, 13.60.Hb \ \ \ MSC: 65c05\\
Keywords: Metropolis Algorithm, Quantum field theory, Monte Carlo 

\end{titlepage}
\section{Introduction}

\input metroIntro.tex

\section{The Metropolis algorithm for a QFT}
\label{Algo}

\input metroAlgo.tex

\section{Examples and performances}
\label{Compu}

\input metroCompu.tex

\section{Conclusions and outlook}
\label{S}

\input metroSumma.tex

\appendix

\input metroApp.tex

\section*{Acknowledgements}

We would like to thank A. Irb\"ack for helpful discussions.
SM is grateful to the UK PPARC for financial support.
This work was supported in part by the TMR Network "QCD and Deep Structure of 
Elementary Particles". HK thanks the Theory Group at RAL for the kind
hospitality extended to him while this work was carried out.

\input error.tex

\end{document}

%% file: metroIntro.tex
Over the past few years, high energy
 particle physics has experienced a tremendous
advance in the number of methods developed to calculate
exactly scattering and decay amplitudes of elementary processes.
Several techniques exist nowadays to evaluate matrix elements (MEs)
for multi-particle events, both analytical and numerical, no matter
the actual number of particles involved.

However, in phenomenological studies, 
the evaluation of the MEs, in terms of the external particle momenta,
is only a part of the whole story. In fact, in order  
to perform realistic analyses, given the finite resolution
of particle detectors, a multi-dimensional integration 
over the whole or some region of 
the phase space available to the final state particles has to 
be performed. Indeed, in most
cases, an analytical evaluation of the integrals become extremely 
complicated, if not impossible,
either because of the complexity of the integrand 
and the large number of dimensions
 involved or because of the presence of cuts in the integration 
region\footnote{Besides, in hadron-hadron and lepton-hadron
collisions, the integration over the Parton Distribution
Functions (PDFs) cannot be performed analytically, as these are
accounted for by programs implementing numerical fits to various
data sets.}.

In such cases, one has to necessarily rely on numerical methods.
Moreover, as the dimension of the integral increases, the number 
of evaluations of the integrand function needed in any generalised 
one-dimensional numerical approach inevitably grows exponentially.
Therefore,  the recourse to Monte Carlo (MC) methods
becomes mandatory: it is well known that the rate of convergence
of MC algorithms is independent of the dimensionality of the integral 
\cite{textbook}.

Naive MC algorithms  are typically based on estimating the
average value of the integrand function by sampling the latter
at discrete points generated according to a uniform  statistical 
(i.e., random) distribution.
However, this approach turns out to be unsatisfactory if the integrand
function strongly {\sl fluctuates} over the integration volume, as it is
the case in many high energy particle processes. Thus, a strategy that
reduces the variance of the integrand ought to be incorporated, in order 
to improve upon the tendency to converge to the correct value
(see Ref.~\cite{variance} for details). 

Two approaches have become popular to this end, known
as {\sl stratified} and {\sl importance sampling}. Whereas 
in the latter case the algorithm concentrates where the integrand
 is largest in magnitude, in the former the
function is primarily sampled where the contribution to the error is largest.
Both techniques suffer however  from a shortcoming: namely, the need
to know {\sl beforehand} the behaviour of the integrand over the 
all integration volume, in order to optimise the implementation.
Unfortunately, it is exactly this knowledge that is often missing in 
particle physics phenomenology.

A successful way of improving in this respect is represented by adaptive 
procedures. These normally involve a division of the original integration
region into a predetermined number of `bins' (word that we
adopt here to signify any partition of the integration volume), with a 
certain number of points in each of these, the latter being at times 
redefined so to perform importance sampling {\sl automatically}. 
A very much used example of such a
technique is the program {\tt VEGAS} \cite{lepage}. 
As a matter of fact,  {\tt VEGAS} also makes use of some 
stratified sampling, in order to improve the convergence in low 
dimensions. Because of its efficiency and accuracy,
this algorithm has eventually come to set the standard in 
many particle physics calculations\footnote{For
{\sl parallel} versions of {\tt VEGAS}, see 
Ref.~\cite{parallel}.}.

The appearance of large fluctuations in the integrand is often 
associated to the
presence of `singularities' in the MEs. In this respect, one can broadly
distinguish between integrable singularities (e.g., resonances) 
inside the phase space and  non-integrable ones (e.g., infrared
emissions)  at its borders. 
Taken separately, they may both be considered as `factorisable'.
In the sense that, provided a suitable choice (or mapping) of the 
integration variables is adopted, then the integrand can be transformed
into a smoother function everywhere over the space space, with the MC
points being generated according to a (suitable)
non-uniform probability density.
In many examples in high energy physics, however, the two kind of poles 
can occur at the same time and,
possibly, there can be more than one of each type: particularly, as the number
of final state particles, their nature and their production channels 
proliferate. 
Under these circumstances, 
the singular structure of the integrand becomes `non-factorisable', in the
sense that there generally exists no change of variables that allows 
even the
adaptive algorithm to sample simultaneously all the singularities
of the integrand in an efficient manner. 
A multi-channel approach can prove useful in such cases \cite{kleiss}.
Here, the actual mapping used to generate a single event is chosen 
randomly, using a predetermined set of probabilities (or weights). 
A combination of the {\tt VEGAS} algorithm with the adaptive multi-channel
sampling of Ref.~\cite{kleiss} has recently been proposed \cite{ohl}.

Another example of adaptive MC algorithm is the  Metropolis technique 
\cite{metroalg}. Widely used in statistical physics (see Ref.~\cite{review} 
for a review), it has nonetheless seen
very little applications to particle physics. These have been mainly confined
to the case of Lattice Gauge Theories (LGTs) \cite{lattice}.
It is our intention here to demonstrate its potential in the context
of a Quantum Field Theory (QFT) of the continuum: in 
evaluating total and 
differential cross sections in multi-particle scattering processes
at high energies. In particular, we will show how a Metropolis
algorithm easily lends itself
to several manipulations, that make it a versatile instrument
in performing such calculations, thus overcoming most of the problems
that we have described, often matching in speed, efficiency
 and accuracy {\tt VEGAS}
itself and outclassing many others of the most commonly used algorithms.

The plan of the paper is as follows. In Sect.~\ref{Algo} we
describe the fundamentals of the Metropolis technique and propose several 
integration methods.
Sect.~\ref{Compu} presents a few numerical examples and comparisons with
other algorithms in the context of some benchmark processes in high energy
particle physics. In Sect.~\ref{S} we summarise our results. Finally,
we will defer to the Appendix the  analytical proof of a condition
required to the algorithm for its convergence in four-momentum phase space.

%% file: metroAlgo.tex
The Metropolis method \cite{metroalg} is a somewhat adaptive MC
integration algorithm which is widely used in numerical statistical mechanics
and LGTs.
We will here demonstrate how it can be modified to evaluate cross 
sections and other observables from perturbative ME calculations 
of scattering and decay processes in a QFT. 
The integration is performed over the four-momentum phase space of the
final state particles. The phase space can be of
arbitrary form, with the only requirement
that the particles are on their mass-shell, i.e., $p_i^2 = m_i^2 
=~{\rm{constant}}$, where $p_i$ and $m_i$ represent the four-momentum
and mass of the $i$-th particle. This restriction is enforced by construction,
in order to give a correct description of the phase space 
associated to the final state
particles. However, intermediate particles, that can appear in a process, are
allowed to be off-shell.

\subsection{Description of the algorithm}
\label{algdesc}
In general, the problem is defined by a phase space and a weight-function  
$w(x)$\footnote{In statistical 
mechanics, $w$ is normally given by an energy-function, a temperature and the 
Boltzmann distribution.}  for every point $x$ in 
phase space. A random walk in the phase space is performed
by starting at an arbitrary initial point $x_0$, and generating new points
$x_i$ by using the weight $w$. 
In the Metropolis algorithm, the sequence $\{ x_i\}$ is generated in a way
which ensures that the points will reach the correct distribution when the 
number of steps (hereafter, $N$) is large. In a QFT, 
the phase space is given by the on-shell 
four-momenta of the outgoing (and possibly incoming) particles, the
weight-function coincides with  the ME, and the 
cross section will correspond to the partition function (defined below) of the
problem. 

In \cite{metroalg} it is shown that in order to reach the correct probability
distribution, $P(x)\propto w(x)dx$, for the sequence, one has to fulfill
two conditions in the stepping procedure $x_i\rightarrow x_{i+1}$.
\begin{itemize}
\item[(a)]{All points in phase space must be reachable with a finite number
of steps.}
\item[(b)]{The condition of `detailed balance' must be fulfilled:
$$
P(x_i)P(x_i\rightarrow x_{i+1})=P(x_{i+1})P(x_{i+1}\rightarrow x_i).
$$
}
\end{itemize}
The normal procedure to satisfy detailed balance (b) is to randomly choose 
a point $\tilde{x}_{i+1}$ with even distribution inside a region $\Omega_i$. 
This new point is then accepted with probability  
$P=\min (1,w(\tilde{x}_{i+1})/w(x_i))$. 
If it is not accepted, the previous point will be put in the sequence once 
more: i.e., 
$x_{i+1}=x_i$. Condition (a) is then satisfied if the overlap of
$\{\Omega_i\}$ can cover the whole of the phase space. 
Choosing the $\Omega_i$'s
to actually be smaller than the phase space will make the Metropolis
algorithm adaptive since, at each step, the 
suggested point $\tilde{x}_{i+1}$ is
likely to be in the region of large weights.

For any observable, ${\cal O}(x)$, the mean can then be calculated as follows:
\begin{equation}
<\!{\cal O}\!>_w\equiv
\frac{\int{{\cal O}(x)w(x)dx}}{\int{w(x)dx}}=\lim_{N\rightarrow\infty}
\frac{\sum_i^N{{\cal O}(x_i)}}{N}.
\end{equation}
The partition function, $Z$, can be calculated by evaluating the average 
of $1/w$ and using the relation:
\begin{equation}
Z\equiv\int w(x)dx=\frac{\int dx}{<\!1/w\!>_w}.
\end{equation}
However, a 
straightforward use of the Metropolis algorithm will in this case be 
inefficient since $1/w$ has large contributions when $w$ is small. The regions
of small $w$ are not visited so often, and one would have large statistical 
fluctuations. For a $w$ that has large variations, it is then 
better to weight the random walk with $w^{1/2}$ instead, and to evaluate $Z$ by:
\begin{equation}
\label{parti1}
Z\equiv\int w(x)dx=\frac{<\!w^{1/2}\!>_{w^{1/2}}}
{<\!w^{-1/2}\!>_{w^{1/2}}}\int dx.
\end{equation}
In this way, the magnitude of the variations in the weight-function are 
effectively halved.
The total volume of phase space, $\int dx$, has to be estimated separately.

In a QFT one is often
interested in evaluating the total cross section other than observables
of the four-momenta. For example,
for processes with two particles in the initial and $n$ in
the final state (scattering reactions), the cross section can be written as
\begin{equation}\label{sigma}
\sigma=\frac{(2\pi)^4}{2p_{tot}^2}\int{
{|\cal{M}|}^2(p_a,p_b,\{p_i\})\prod_{i=1}^n{\left[
\frac{d^4p_i}{(2\pi)^3}
\delta^+\!\left(p_i^2-m_i^2\right)\right]}
\delta\!\left(p_{tot}-\sum{p_i}\right),}
\end{equation}
where $p_{tot}$ is the total four-momentum, $p_a$ and $p_b$ are the incoming
momenta, $\{p_i\}$ are the $n$ final state momenta, with $m_i$ their respective
masses, and 
$|{\cal M}|^2$ the ME associated to the reaction considered. 
By choosing the weight-function and the phase space to be 
\begin{eqnarray}
w&=&\frac{(2\pi)^{4-3n}}{2p_{tot}^2}{|\cal{M}|}^2(p_a,p_b,\{p_i\}),
 \label{weight} \\
\label{PS}
dx&=&\prod_{i=1}^n{\left[
d^4p_i
\delta^+\!\left(p_i^2-m_i^2\right)\right]}
\delta\!\left(p_{tot}-\sum{p_i}\right),
\end{eqnarray}
respectively, we can use the Metropolis algorithm to perform
the integration.

In order to do so, one has to guarantee that 
the stepping procedure in the Lorentz-invariant
phase space (\ref{PS}) can be done in a way that satisfies the two conditions
(a) and (b) above. We now describe how this can be achieved:
first comes an illustration (points (i) to (iv)) of how the
suggested point in phase space, $\tilde{x}_{i+1}$, can be
generated; afterwards, we show that this procedure complies with  (a)
and (b). 

The proposed point in phase space, $\tilde{x}_{i+1}$, can be
generated  in the following way.
\begin{itemize}
\item[(i)]{Choose two of the final state particles randomly.}
\item[(ii)]{Boost them into their centre-of-mass (CM) frame.}
\item[(iii)]{In that frame, rotate them randomly with an even distribution in 
$4\pi$.}
\item[(iv)]{Boost them back into the original frame.}
\end{itemize}

In order to demonstrate
 that this procedure will satisfy conditions (a) and (b), we
will make use of the Lorentz-invariance of $dx$ in eq. (\ref{PS}). To make sure
that (b) is fulfilled, we have to show that the suggested point is generated
with an even distribution inside the region $\Omega$, which is reachable in one
step. But it is clear that
this is true in the CM frame of the two particles chosen in (i). Therefore,
and because of Lorentz-invariance, it is true in any frame.

To demonstrate
 that also condition (a) is fulfilled, we have to show that from any
point in phase space, $\{p_i\}$, one can reach any other point, $\{p_i'\}$,
by a finite number of steps. This can actually be achieved in, at most,
$n-1$ steps. Let $\Delta p_i$ be the shift of the $i$-th four-momentum,
$p_i'=p_i+ \Delta p_i$. Choose an arbitrary particle, say $p_n$. 
For each of the $n-1$ other momenta, 
$\Delta p_i$ can be transferred to $p_n$ in one step. This is clear since
$p_i+p_n$ is conserved. In the CM frame of $p_i$ and $p_n$, the whole 
surface of conserved $p_i+p_n$ and the other momenta fixed can be reached in 
one step. So is in any frame:
any $\Delta p_i$ which conserves $p_i+p_n$ can be transferred in one step.
By doing this for all the $n-1$ other particles, we have reached the point
$\{p_i'\}$.

In some cases, one might want to introduce cuts in the
integration volume\footnote{Alternatively, the ME might be 
identically zero inside a finite region of it.}. That is, instead
of integrating over the whole of phase space, $PS$, one might want to integrate
inside a reduced region, $PS_{\rm{red}}$. 
In that case, one has to make sure
that condition (a) is fulfilled also over
 $PS_{\rm{red}}$. In the Appendix
we show that this is the case if $PS_{\rm{red}}$ is a connected region. If this
is not true, then $PS_{\rm{red}}$ has to be separated
into parts, $PS_{{\rm{red}},i}$, each of which is 
connected, and the integration be done for each of these separately.

Finally, in some cases, it can be desirable to integrate a ME (or just
a term of it, in which case we use the notation 
${\cal T}$) which is 
negative over some parts of the phase space. If so, ${\cal T}$ can not 
directly be used as the weight-function $w$. 
Instead, one can use its absolute value, and keep track of how often 
${\cal T}$  is
negative. With $w=|{\cal T}|$, the partition function is in this case given 
by:
\begin{equation}
\label{parti3}
Z=\frac{<\!{\rm sign}({\cal T})~w^{1/2}\!>_{w^{1/2}}}
{<\!w^{-1/2}\!>_{w^{1/2}}}\int dx.
\end{equation}

In summary, the average of an observable $\cal{O}$, given by a
function ${\cal O}(\{p_{i}\})$, can be calculated as follows.
\begin{itemize}
\item{Choose a suitable starting point $\{p_{i0}\}$.}
\item{Generate a new point $\{\tilde{p}_{i1}\}$ as described above.}
\item{Accept this point with the probability
$$
P\left(\{p_{i1}\}\!=\!\{\tilde{p}_{i1}\}\right)=
\min\left(1,\frac{w(\{\tilde{p}_{i1}\})}{w(\{p_{i0}\})}\right),
$$
with $w$ given by the ME (eq. (\ref{weight})).}
\item{If not accepted, use the previous point: $\{p_{i1}\}=\{p_{i0}\}$.}
\item{Store the value of $\cal{O}$: 
${\cal O}_{\rm{sum}}={\cal O}_{\rm{sum}}+{\cal O}(\{p_{i1}\})$.}
\item{Repeat the four last steps until the desired level of convergence is
reached ($N$ steps).}
\item{Finally, take the average: $<\!{\cal O}\!>\approx 
\frac{{\cal O}_{\mathrm{sum}}}{N}$.}
\end{itemize}

\subsection{Time consumption and error estimates}
In most of the (sub)processes that we have
integrated with the Metropolis algorithm,
the, by far, most time-consuming tasks have been the calls to the
ME function. Our 
strategy to reduce the CPU-time has then been to, as far as possible, reduce
the error for a fixed number of calls to this function.

The algorithm requires many boosts and rotations of four-vectors. For 1 million
steps through phase space, e.g., this takes around 30 seconds on a 350 MHz
Pentium-II processor, on a LINUX platform. Calling the ME the same number of 
times on the same machine is often a matter of hours. Thus, the stepping
procedure has not been optimised. Instead, we have tried different ways of
refining the interface between the algorithm itself and its use of the ME. 

A complication in evaluating the statistical error in a Metropolis-based
integration
is that the generated points in phase space are correlated. The correlation
length effectively reduces the number of statistically independent 
data points.  
We have handled this flaw by collecting the data into a number of `blocks'. The
average in each block is then taken and used as a new, 
statistically independent
data point. The size of the blocks must be made larger than the correlation
length, while the number of blocks must be large enough, so that the variance
evaluated from the averages can be trusted.
The normal procedure to check that the correlation length has been 
reached is to
divide each blocks into smaller parts, and check that single 
block-parts inside one
block are not more correlated than the block-parts from different blocks,
see Ref.~\cite{blockmeth}.

\subsection{Some methods of refinement}
In this Subsection, we will describe some methods of refinement which are
devoted to reduce the error in the integration for a certain number of calls
to the ME. These methods will in general increase the CPU-time of the
algorithm significantly, but, as mentioned above, the overall CPU-consumption
will increase marginally, while improving the efficiency of the algorithm
in various respects. 

\subsubsection{The parallel integration method}
\label{parall}
In realistic problems, the ME can present 
large variations, often due to resonances arising from
intermediate particles. In simple terms,
large variations make it hard to move around in
phase space. The correlation length in the 
Metropolis steps will be large and many
calls to the ME\footnote{Hereafter, denoted by the short-hand notation
$|{\cal M}|^2(x)\equiv|{\cal M}|^2(p_a,p_b,\{p_i\})$, where $x$ represents
the phase space point $\{p_i\}$ of eq.~(\ref{sigma}).}
will be required in order to get independent
data points. We will here describe a method, the `parallel integration
method', which enables one to move large distances in phase space,
without using the exact (time consuming) ME function.

First, one can introduce an approximate ME,
$|\widetilde{{\cal M}}|^2(x)$. 
This should be a simple function which is quick to call, and
which is a good approximation in the regions where $|{\cal M}|^2(x)$ has large
contributions. It could, e.g., be a product of just the resonant propagators 
in $|{\cal M}|^2(x)$.
Then the configuration space is enlarged with an extra, discrete 
parameter, $s$, which can
assume the values $0$ or $1$. A weight-function $w$ is constructed in the
enlarged configuration space, such that it returns $|{\cal M}|^2(x)$ if $s=1$ 
and
$|\widetilde {{\cal M}}|^2(x)$ if $s=0$. Finally, the Metropolis-step is 
modified so that,
occasionally, instead of suggesting a new set of four-momenta, a new value of
$s$ is suggested. Data points are to be taken only when $s=1$. The
distributions of points in the part with $s=1$ (of our enlarged
configuration-space) will be exactly as in the original phase space but with
only $|{\cal M}|^2(x)$ as the weight function. We know this since, in general, the
Metropolis algorithm ensures that the generated points are distributed
according to the weight-function, and this is of course true also for
individual parts of the configuration space. 

The gain of this method is that, during the steps in
phase space when $s=0$, one can reach distant points, without calling the 
slow, exact $|{\cal M}|^2(x)$. At will, one can choose to make it 
more probable that
$s=0$, so that $|{\cal M}|^2(x)$ is used as seldom as possible. For
example, this can be done by
introducing two integers, $N_0$ and $N_1$. If $s=i$, we attempt to switch
the value of $s$ after $N_i$ steps. We can then choose, e.g., $N_0=100$ and
$N_1=1$. The optimal value of $N_0/N_1$ will be given by how faster the
approximate ME function actually is in comparison to the exact one.

Let us summarise the parallel integration method.
\begin{itemize}
\item{Find a crude and quick approximation,
 $|\widetilde{{\cal M}}|^2(x)$, to the original
    ME, $|{\cal M}|^2(x)$.}
\item{Enlarge the phase space by introducing the discrete
parameter $s=\{0,1\}$.}
\item{Construct the weight function
$$
w(x,s)=
\left\{
\begin{array}[c]{c}
|{\cal M}|^2(x) \ \ \ {\rm{if}} \ \ s=1,\\
|\widetilde{{\cal M}}|^2(x) \ \ \ {\rm{if}} \ \ s=0.\\
\end{array}
\right .
$$
}
\item{Choose an arbitrary starting point $x_0$ in phase space and set
    $s=0$\footnote{Also arbitrary, but $s=0$ is recommended.}.}
\item{Perform $N_0$ Metropolis steps with the weight 
$w(x,0)=|\widetilde{{\cal M}}|^2(x)$,
    and without collecting the value of the observable ${\cal O}$.}
\item{Switch the value of $s$ with probability
$$
P=\min\left(1,\frac{w(x,1)}{w(x,0)}\right)=
\min\left(1,\frac{|{\cal M}|^2(x)}{|\widetilde{{\cal M}}|^2(x)}\right).
$$
}
\item{If the switch is accepted, store the value of the observable.}
\item{The last three points are repeated in the following way until desired
    convergence is reached:
\begin{itemize}
\item{If $s=i$, perform $N_i$ Metropolis steps and, only if $s=1$, store the 
    value of $\cal O$ after each step.}
\item{Switch the value of $s$ with the probability
$$
P=\min\left(1,\frac{w(x,s')}{w(x,s)}\right),
$$
where $s'=1$ if $s=0$ and $s'=0$ if $s=1$.}
\item{Store the value of $\cal O$, if the new $s=1$.}
\end{itemize}
}
\item{Take the average of the observable.}
\end{itemize}

With this method, the total cross-section is calculable in a more effective
way, than described before. This is the case if the cross-section of the 
approximate ME is already known with high accuracy. By counting the number of 
times, $r_i$, that $s=i$ after a switch, 
one gets the relative magnitudes of the
cross-sections from the fraction of the $r_i$'s. If the exact cross-section is
denoted by $\sigma$, and the approximate one by $\widetilde{\sigma}$, one has
that 
\begin{equation}
\sigma=\widetilde{\sigma}\frac{r_1}{r_0}.
\end{equation}

\subsubsection{The variable energy method}
\label{ene}
In some cases, one is interested in doing the integration with a
dynamically variable total energy, $E_{\mathrm{tot}}\ne~{\mathrm{constant}}$. 
This might be the case,
e.g., if one is interested in the cross section as a function 
of $E_{\mathrm{tot}}$ in a
certain range, or if the two incoming particles can have varying energy, e.g.,
depending on some given PDF. To this end,
we describe here the `variable
energy method', where the configuration space is extended 
by adding $E_{\mathrm{tot}}$
as a dynamical parameter. Allowing for a variable energy can actually make the
integration more effective, for the following reason: it is often the case
that the variations of the ME are larger for larger energies. The
phase space walk is thus easier at low energies. If one is interested in some
observables to be calculated at large $E_{\mathrm{tot}}$, it can be easier 
to reach farther
in phase space by, say, taking a round tour into the lower energy regions. 
This idea has been explored in calculations in statistical mechanics, where
often is the temperature to be used as a dynamical variable 
\cite{dyntemp,dynvar}.

One proceeds as follows.
Let a point in the extended configuration space be denoted by 
$x=(\{p_i\},E_{\mathrm{tot}})$, where $E_{\mathrm{tot}}$ has been 
written explicitly,
though  it is actually computed from the four-momenta. 
In order to be able to dynamically
take steps to other energies, we introduce a one-to-one mapping 
$\{\tilde{p}_i\}=K(\{p_i\},\tilde{E}_{\mathrm{tot}})$, which gives a new 
point, 
$(\{\tilde{p}_i\},\tilde{E}_{\mathrm{tot}})$, given the previous one, 
$(\{p_i\},E_{\mathrm{tot}})$, and the new energy as well, 
$\tilde{E}_{\mathrm{tot}}$. 
We also need a
phase space weight,
$\rho[(\{p_i\},E_{\mathrm{tot}})$ $\rightarrow$
$(\{\tilde{p}_i\},\tilde{E}_{\mathrm{tot}})]$, which is
evaluated from phase space densities at the different energies (as described
below). In the Metropolis path, one can alternatively attempt to change the
energy or just update the configuration. A step which can 
change $E_{\mathrm{tot}}$ is 
then taken in the following way.
\begin{itemize}
\item{Suggest a new energy, $\tilde{E}_{\mathrm{tot}}$, chosen with even distribution
    in a certain region.}
\item{Find the corresponding point in phase space,
$\{\tilde{p}_i\}=K(\{p_i\},\tilde{E}_{\mathrm{tot}})$.}
\item{Accept this with probability
$$
P=\min\left(
  1,\rho[(\{p_i\},E_{\mathrm{tot}})\rightarrow(\{\tilde{p}_i\},\tilde{E}_{\mathrm{tot}})]
\frac{|{\cal M}|^2(\{\tilde{p}_i\})}{|{\cal M}|^2(\{p_i\})}\right).
$$}
\end{itemize}

How  the mapping $K$ can be defined and how the corresponding
phase space weight $\rho$ is calculated is the next step. In order to describe
how this can be achieved, we have to digress briefly, by giving a general
description of the Lorentz phase space, defined in eq. (\ref{PS}) (for an
overview, see, e.g., Ref. \cite{lorph}).

Let $V_n(\{m_i\},s)$ denote the total volume of the $n$-particle phase space
with masses $\{m_i\}$ at the squared invariant energy $s$. For $n=2$, we
have
\begin{equation}
V_2(m_1^2,m_2^2,s)=\frac{\pi}{2s}\sqrt{\lambda(m_1^2,m_2^2,s)},
\end{equation}
with $\lambda(a,b,c)=a^2+b^2+c^2-2ab-2bc-2ca$. In the CM frame, the
magnitude of the two outgoing momenta is given by
\begin{equation}
\label{momfunc}
P(m_1^2,m_2^2,s)=\frac{1}{2\sqrt{s}}\sqrt{\lambda(m_1^2,m_2^2,s).}
\end{equation}
The volume for $n$ particles
can then be calculated recursively:
\begin{equation}
\label{volume}
V_n(\{m_i\}_n,s)=\int_{s_0}^{s_1}ds'
{V_2(s',m_n^2,s)V_{n-1}(\{m_i\}_{n-1},s'),}
\end{equation} 
where the integration range is restricted
 between $s_0=\left(\sum_1^{n-1}{m_i}\right)^2$
and $s_1=\left(\sqrt{s}-m_n\right)^2$.
The mapping $K$, from the energy $E_{\mathrm{tot}}$ 
to ${\tilde{E}}_{\mathrm{tot}}=E_{\mathrm{tot}}+\Delta E$, can be
defined by changing the momentum of the $n$-th particle and boosting the 
others, so that the same CM frame is kept. In case this is actually
possible, the back-to-back momentum of the $n$-th particle and the $n-1$
particle system, in the CM frame, is given by 
$P(s',m_n^2,{\tilde{E}}_{\mathrm{tot}}^2)$, with
the momentum-function $P$ as in eq. (\ref{momfunc}).

When the change in energy can be done in this way (i.e.,
provided  $\lambda\!>\!0$),
$\rho$ is given by the ratio of the 2-particle volume  for the two energies. 
This is the case since the value of $s'$ is preserved and the $n-1$ particle 
volume in the integrand in eq. (\ref{volume}) is not affected. 
Thus, the $\rho$ expression 
corresponding to our mapping $K$ is given by
$$
\rho(E_{\mathrm{tot}}\rightarrow{\tilde{E}}_{\mathrm{tot}})=
\frac{V_2(s',m_n^2,{\tilde{E}}_{\mathrm{tot}}^2)}
{V_2(s',m_n^2,E_{\mathrm{tot}}^2)}
\theta(\lambda(m_1^2,m_2^2,s)),
$$
where the $\theta$-function returns a zero when $\lambda\!<\!0$ 
(corresponding
to the case in which $s'$ is out of the integration range of
 eq. (\ref{volume})), and 1 otherwise. We
stress again that the choice of the mapping $K$ is not unique. We have 
illustrated here 
a simple method, however, it is possible that other, more sophisticated
mappings can be chosen to make the whole procedure more efficient.

\subsubsection{Variable energy and parallel integration}
\label{varpar}
The two methods described above can be used simultaneously, as follows. 
Choose an energy
interval $IE$, where the integration is to be performed, and a fixed energy,
$E_p$, to be used only for the approximate ME 
function $|\widetilde{\cal M}|^2(x)$. Also
define a smaller interval $IE_W\subset IE$, which will be the `window' where we
switch between the exact and approximate MEs. The need for such a region is
induced by the fact that the cross section often varies a lot with energy. 
The window should be
put around the region in energy where the cross section is expected to be
large. It is also
recommendable to have the fixed energy, $E_p$, inside this window. The
algorithm then goes as for the parallel integration method
 except that, when the exact ME is to be used, one allows also 
for energy updates. 
We here describe how the cross section as a function
of energy is evaluated, assuming that the cross section
for the approximate ME, $\tilde{\sigma}(E_p)$, is known at one energy, $E_p$. 

The stepping procedure is done as sketched below.
\begin{itemize}
\item{Choose the energy ranges $IE\supset IE_W$, and the point $E_p\in
    IE_W$: $IE$ is the range we are interested in, $IE_W$ should be chosen in
    the region of large cross sections and $E_p$ is the point where
    $\tilde{\sigma}$ is known. Let, as before, $E_{\mathrm{tot}}$ 
    denote the latest accepted energy used for the exact ME. The starting value
    could, e.g., be $E_{\mathrm{tot}}=E_p$.}
\item{Introduce the discrete parameter, $s=\{0,1\}$, and the two update 
    periods, $N_0$
    and $N_1$, as before.}
\item{Start with $s=0$ at the energy $E_p$ and perform $N_0$ ordinary
    Metropolis steps with the approximate ME.}
\item{Attempt a switch of the value of $s$ by proposing an energy step
    $E_p\rightarrow E_{\mathrm{tot}}$, as described in Subsect. \ref{ene}, 
    with the new 
    weight given by the exact ME.}
\item{Whenever $s=1$ is accepted, an energy-varying Metropolis path is taken
    in the region $IE$. Whenever $N_1$ points inside the window $IE_W$
    have been chosen, a new switch of $s$ is attempted to the point $E_p$
    and with the approximate ME as the new weight.}
\end{itemize}

In order to get an estimate for the total cross section
for the exact ME, $\sigma(E)$,  one needs to
store the distribution of energies generated when $s=1$. This distribution is
then normalised by means of  $\tilde{\sigma}(E_p)$,
together with an evaluation of how often $s=1$. That is,
 proceed as follows.
\begin{itemize}
\item{Whenever $s=1$ and an energy step inside $IE$ is taken, put the
    new\footnote{
      This could be the same energy as before, in case the suggested energy is
    not accepted.}
    energy into a distribution, $f(E)$.}
\item{Evaluate the number of times, $r_i$, that the value of $s=i$, after a
    switch attempt.}
\item{Do this until desired convergence of the distribution, $f(E)$, and the
    numbers, $r_i$, is reached.}
\end{itemize}
The cross section then becomes, $\sigma(E)=Nf(E)$ with the normalisation
factor $N=\tilde{\sigma}(E_p)r_1/\bar{f}r_0$, where $\bar{f}$ is the average
of $f(E)$ inside the window $IE_W$.


%% file: metroCompu.tex
\def\Ord{\buildrel{\scriptscriptstyle <}\over{\scriptscriptstyle\sim}}
\def\OOrd{\buildrel{\scriptscriptstyle
    >}\over{\scriptscriptstyle\sim}}
\def\NMC{\ifmmode{N_{\mathrm{MC}}}
             \else{$N_{\mathrm{MC}}$}\fi}
\def\NACC{\ifmmode{N_{\mathrm{ACC}}}
             \else{$N_{\mathrm{ACC}}$}\fi}
\def\NGEN{\ifmmode{N_{\mathrm{GEN}}}
             \else{$N_{\mathrm{GEN}}$}\fi}

In this Section we report  about some numerical results obtained by using the
Metropolis algorithm and compare them to the outputs of other 
integration programs widely used in particle physics calculations.
In Subsect.~\ref{photons} we study multi-photon production in 
electron-positron annihilations, as an example of the ability of
the various algorithms to deal with the problem of singularities
associated with an increasing number of infrared emissions of
different topology and over an increasingly reduced phase space. 
The benchmark process
considered in Subsect.~\ref{eebbww} is the production of $b$-quark
and $W^\pm$-boson pairs at future 
lepton colliders, in order to test
the performances of the algorithms in presence of massive particles and
divergent/resonant invariant mass poles,
the latter further occurring in a multi-channel environment where
use of negative weights is made. Finally, in Subsect.~\ref{hadron}, we 
will focus
our attention to the case of some radiative top-antitop events as
produced in hadron-hadron collisions, leading to eight-parton
final states, the latter integrated over a reduced phase space:
here, we account for the behaviour of the integrator as induced
by a non-fixed partonic energy, a large number of particles in the
final state and the enforcement of experimental cuts through 
$\theta$-functions in the integrand.

We have carried out our tests on several machines, without finding
any significant dependence of the relative algorithm performances
upon a particular choice among them. For reference, we list here
the platforms used: $\alpha$-DECstations 3000 Model 300 and 400
and $\alpha$-Servers 1000A 5/300 and 1000 4/200 
(running both VMS and OSF operating systems), 
UNIX Hewlett-Packard Workstations
type A 9000/712 and A 9000/782 and the LINUX system already mentioned.

\subsection{Multiple infrared radiation}
\label{photons}

The reference paper for this Subsection is \cite{WJS}.
There, it was studied the tree-level process $e^+e^-\to n~{\gamma}$
 (with massless electrons/positrons), where
$\gamma$ refers to a photon with $n$ taken up to 7,
 and where the relevant MEs were presented analytically.
In our forthcoming tests, we have made use of the expression for the
latter given in eq. (17) of \cite{WJS}. Notice that 
such a formula is exact for $n<4$ only, whereas it is an approximation for 
$n\ge4$.
However, being much faster than the exact ME, eq. (16) of that paper,
while retaining its main dynamic features, it is much more useful to our 
purposes.

The challenge here is to integrate a ME that diverges when 
a photon becomes either soft or collinear (with one of the initial
state fermions). To render the results
 finite, while still allowing for the infrared
behaviour at the edges of the (reduced)  phase space, we impose the following
cuts (same as in Ref.~\cite{WJS}):
\begin{equation}\label{photoncuts} 
E_\gamma>5~{\mathrm{GeV}},\qquad\qquad\qquad\cos\theta_\gamma<0.9,
\end{equation}
on the energy and (cosine of the) polar angle of each final state
photon. For comparison purposes, we also adopt the same
CM energy used in \cite{WJS}, that is, $\sqrt s=100$ GeV.
However, we will increase here the number $n$ of photons produced to 9.

The algorithms used for this example, other than the Metropolis one,
are {\tt VEGAS}, {\tt RAMBO} and the {\sc NAGLIB} subroutines {\tt D01EAF}
and {\tt D01GCF}. Of {\tt VEGAS}, we have already given a description.
{\tt{RAMBO}} \cite{RAMBO} is not exactly an integrator, though it can be
used in some instances in such a fashion. Rather, it is a 
multi-particle phase space generator, as -- given the total CM
energy and the number of particles required with their masses -- it
produces a set of four-momenta and the phase space weight associated
with that configuration. However, when the integrand function does not present
sharp peaks (as it is the case here), it can be used to estimate the
total cross section and its standard deviation simply as an 
arithmetic average and through a standard quadrature formula, respectively.
Indeed, this is the algorithm that was successfully employed in 
Ref.~\cite{WJS}. The subroutines {\tt D01EAF}
and {\tt D01GCF} are part of the {\sc NAGLIB} library
\cite{NAGLIB}. They both are multi-dimensional adaptive quadrature
integrators, the first over an  hyper-rectangle  whereas the 
second on a general product region. {\tt D01EAF} makes use of the
algorithm described in Ref.~\cite{D01EAF}, whereas for {\tt D01GCF} 
one should refer to \cite{D01GCF}. Their usage
and characteristics  are well introduced in Ref.~\cite{NAGLIB}, so
we do not dwell any longer on them.
The Metropolis implementation used here is the one described in
Subsect. \ref{algdesc}, making use of the more effective formula in
eq. (\ref{parti1}).
The reduced phase space
volume (after cuts), i.e., $\int{dx}$ in (\ref{parti1}),
was calculated numerically but with insignificantly small errors.
Alternatively, one can assume it to be given as an exact input, if
the phase space integration can be performed analytically.

In this test, we have proceeded as follows. For a start, we have
fixed the number
of MC points generated to carry out the integration,
\NMC, to be approximately 
$10^6$ (including 
those eventually rejected because of the cuts) for all
algorithms and irrespectively of the number of photons generated.
One may consider this condition as a prototype of what inevitably occurs
in numerical computations, when one can only dispose of a finite amount of 
CPU (here, corresponding to the time needed to evaluate
$10^6$ times the integrand function, as it would actually happen
if all momentum configurations generated were accepted)\footnote{In fact, 
here the expression of the $n$-photon ME is considerably simple that the time
spent in evaluating it has little impact on the total one employed by 
the all integration procedure, irrespectively of the value of $n$.}.
Notice that we have introduced the cuts through theta functions in energy and
angles, while maintaining as upper and lower limits of the integration
variables chosen those needed to cover the all of the original (massless) 
$n$-particle phase space\footnote{For consistency, we have customised
the choice of the latter to be the same invariant masses and angles for
{\tt VEGAS}, {\tt D01EAF} and {\tt D01GCF}. As for {\tt RAMBO}, there
is no need to provide the integration variables and corresponding
Jacobian factors, as this is done automatically by the program: see
Ref. \cite{RAMBO} for specific details.}. 
Under these conditions, what one would expect from an optimal algorithm is
both the tendency of promptly adapting itself to an increasingly reduced
integration volume -- as $n$ gets larger -- (thus minimising the loss 
of MC points through the cuts) and the ability to efficiently dispose
of those points that survive the kinematic 
constraints (by minimising the error of the integration).
A measure of the former is the ratio between accepted and generated MC
points, whereas for the latter of the percentage spread of the errors 
about the central values obtained. We will see that the estimates
 will roughly be consistent with each other 
for all algorithms up to $n=8$, while
 some of the programs will fail to converge for
$n=9$.
Then, for those programs that manifest the ability to  converge to the 
correct value of the cross sections with good accuracy up to $n=9$, 
we have increased
$N_{\mathrm{MC}}$ by ten and hundred times, 
this way further checking 
the rate of convergence of the algorithms when, ideally, an 
infinite CPU (i.e., number of MC points) 
is made available. For sake of illustration, we have reproduced
in this case the $e^+e^-\to 7~\gamma$ cross section.

\begin{figure}[htb]
\begin{center}
~\epsfig{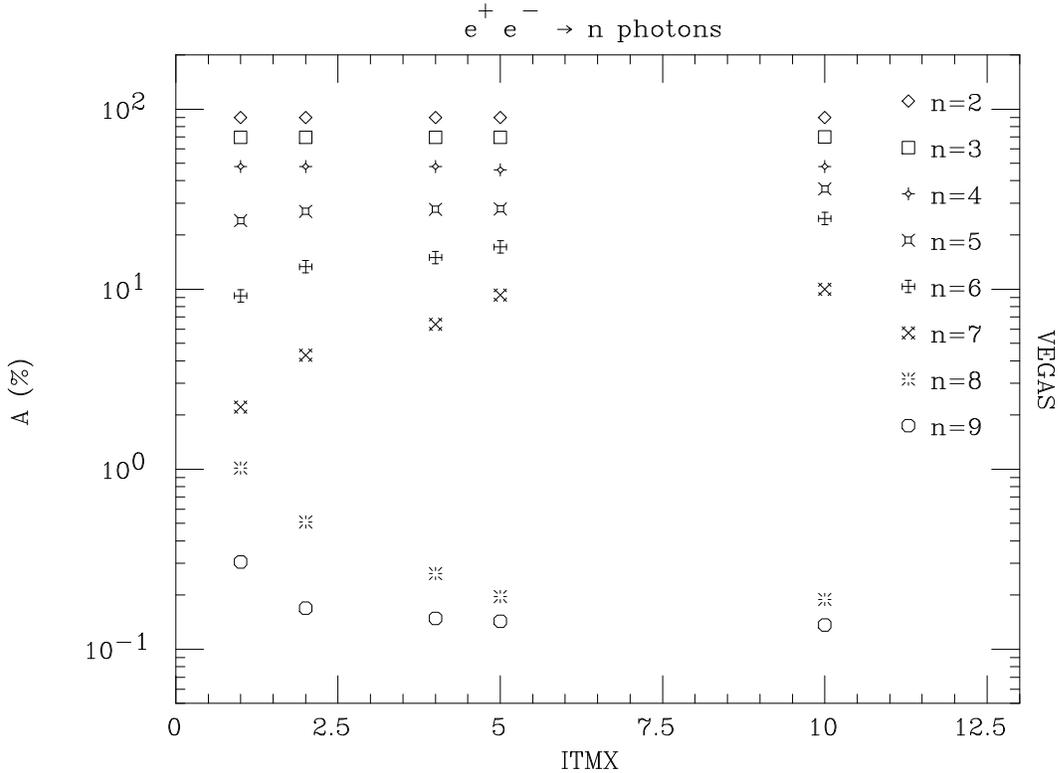}
\caption{Acceptance rates, $A$, defined as the ratio of accepted over
generated MC points (times 100), for the {\tt VEGAS} algorithm in evaluating
the total cross section for $e^+e^-\to n~\gamma$, with $n=1, ... 9$,
for five possible choices of ITMX and NCALL, such that their product
is approximately equal to $\NMC=10^6$.}
\label{fig:acceptance}
\end{center}
\end{figure}

Before proceeding to compare the algorithms, one should recall that
in {\tt VEGAS} there exist two parameters, {\rm{NCALL}} and {\rm ITMX}, 
that determine the actual number of points generated. The first is the
(approximate) number of integrand evaluations per iteration whereas
the second is the maximum number of the latter\footnote{As we set 
the required accuracy to be negative, i.e., ${\rm{ACC}}<0$, all of them are
performed.}: to change one or the other affects the overall performance of the
algorithm in various respects \cite{lepage}. Thus, as a preliminary exercise, 
we have run {\tt VEGAS} varying these two parameters: e.g., by taking 
${\rm{ITMX}}=1, 2, 4, 5, 10$ 
and, consequently, ${\rm{NCALL}}=\NMC/{\rm{ITMX}}$,
being $\NMC=10^6$. The outcome is presented in Fig.~\ref{fig:acceptance},
where we show the acceptance rate, $A$. 
There, one can first notice the indifference
of {\tt VEGAS} to the choice of NCALL and ITMX for $n\le4$, that is,
if the dimension of the integral\footnote{Note that
we have mapped the $n$-photon phase space in such a way the the
azimuthal angle around the electron/positron beam direction is one of the
integration variables. Being the cross section independent of
this variable, the phase space integral can be reduced by one dimension
and simply multiplied  by $2\pi$. Of course, four 
of the initial $n^3$ dimensions of the phase space 
integral are removed by the $\delta$-functions associated to
the four-momentum conservation between the initial and final states.}, 
NDIM, is below 7. For $n\ge5$, or equivalently ${\rm{NDIM}}\ge10$, 
if one increases ITMX, the adaptability worsen for high dimensionalities
while it improves for low ones. As a compromise between the two
tendencies, we will adopt for the reminder of this test the choice
${\rm{ITMX}}=4$ and 
${\rm{NCALL}}=\NMC/{\rm{ITMX}}$ 
for any $n$. In the case a {\tt VEGAS} iteration
fails to find points above the cuts (as it can happen for very large $n$),
thus yielding a zero, its contribution to the total estimate
of the cross section is discarded altogether.

\begin{table}[htb]
\begin{center}
\begin{tabular}{|c||c|c|c||c|}
\hline
\multicolumn{5}{|c|}
{\rule[0cm]{0cm}{0cm}
$\sigma (e^+e^-\to n~\gamma)$ (nb) at $\sqrt s=100$ GeV}
\\ \hline\hline
$n$ & Metropolis & {\tt VEGAS} & {\tt RAMBO} & $\;$ \\ \hline
2                    &  
$2.66782\pm0.0029$   & 
$2.664973\pm0.000050$& 
$2.66114\pm0.0023$   & 
$\times10^{-2}$      \\ \hline
3                    &  
$6.34056\pm0.029$    & 
$6.26009\pm0.017$    & 
$6.26585\pm0.014$    & 
$\times10^{-4}$      \\ \hline
4                    &  
$6.63656\pm0.036$    & 
$6.63202\pm0.050$    & 
$6.67522\pm0.025$    & 
$\times10^{-6}$      \\ \hline
5                    &  
$4.0304\pm0.034$    & 
$3.94629\pm0.054$    & 
$3.99786\pm0.023$    & 
$\times10^{-8}$      \\ \hline
6                    & 
$1.49256\pm0.013$    & 
$1.48609\pm0.048$    & 
$1.47493\pm0.010$    & 
$\times10^{-10}$     \\ \hline
7                    &  
$3.72003\pm0.037$    & 
$3.73682\pm0.26$     & 
$3.63071\pm0.040$    & 
$\times10^{-13}$     \\ \hline
8                    &  
$5.92104\pm0.072$    & 
$5.51518\pm0.51$     & 
$5.83508\pm0.077$    & 
$\times10^{-16}$     \\ \hline
9                    &  
$6.43459\pm0.079$     & 
$5.62617\pm1.3$      & 
$6.57829\pm0.21$     & 
$\times10^{-19}$     \\ \hline
\end{tabular}
\begin{tabular}{|c||c|c||c|}
\hline
$n$ & {\tt D01EAF} & {\tt D01GCF} & $\;$ \\ \hline
2                    &  
$2.66510\pm0.0019$   & 
$2.66493\pm0.000010$ & 
$\times10^{-2}$      \\ \hline
3                    &  
$6.19822\pm0.29$     & 
$6.15469\pm0.031$    & 
$\times10^{-4}$      \\ \hline
4                    & 
$6.47451\pm4.1$      & 
$6.60557\pm0.13$     & 
$\times10^{-6}$      \\ \hline
5                    & 
$5.13410\pm4.8$      & 
$4.11628\pm0.45$     & 
$\times10^{-8}$      \\ \hline
6                    & 
$11.2745\pm42.$      & 
$1.44837\pm0.21$     & 
$\times10^{-10}$     \\ \hline
7                    & 
$0.776428\pm33.$     & 
$2.87767\pm0.64$     & 
$\times10^{-13}$     \\ \hline
8                    & 
$3.56037\pm385.$     & 
$4.93633\pm2.2$      & 
$\times10^{-16}$     \\ \hline
9                    & 
$59.0009\pm866.$     & 
cannot compute       & 
$\times10^{-19}$     \\ \hline\hline
\multicolumn{4}{|c|}
{\rule[0cm]{0cm}{0cm}
$E_\gamma>5~{\mathrm{GeV}} \qquad\qquad\qquad\cos\theta_\gamma<0.9$}
\\ \hline

\end{tabular}
\caption{The cross sections and relative errors for $e^+e^-\to n~\gamma$
at $\sqrt s=100$ GeV,
with $n=1, ... 9$, as obtained from the five algorithms documented in the 
text. For {\tt VEGAS}, the setup ${\rm{ITMX}}=4$ and ${\rm{NCALL}}=250000$
was adopted. Several values of  MINCLS (see Ref.~\cite{NAGLIB})
were used  for {\tt D01EAF}, but
no significant improvement was found compared to the data reported. }
\label{tab:photons}
\end{center}
\end{table}

In Tab.~\ref{tab:photons} we present, as function of $n$, the central values
and the associated errors produced by the five integrators considered
in evaluating the cross sections for $e^+e^-\to n~\gamma$
(compare to Tab.~1 of Ref.~\cite{WJS} for $n\le7$). ({\tt D01GCF} cannot 
compute integrals with more then 20 dimension, so that the cross
section corresponding to $n=9$ does not appear in the table.)
A first obvious result (apart from the shortcomings of {\tt D01EAF}
as $n$ increases) is that whereas {\tt VEGAS} performs undoubtedly
better than Metropolis for small $n$, say, below 4, if $n\ge4$, 
Metropolis yields a much more accurate answer. Even {\tt RAMBO}, a
non-adaptive algorithm, excels over {\tt VEGAS} for large photon
multiplicities, though, for $n=9$, not as well as Metropolis. The error from
{\tt D01GCF} is significantly larger than that for the other algorithms (except
{\tt D01EAF}) for $n\ge4$, whereas for smaller values it almost achieves 
the accuracy of {\tt VEGAS}.

\begin{figure}[htb]
\begin{center}
~\epsfig{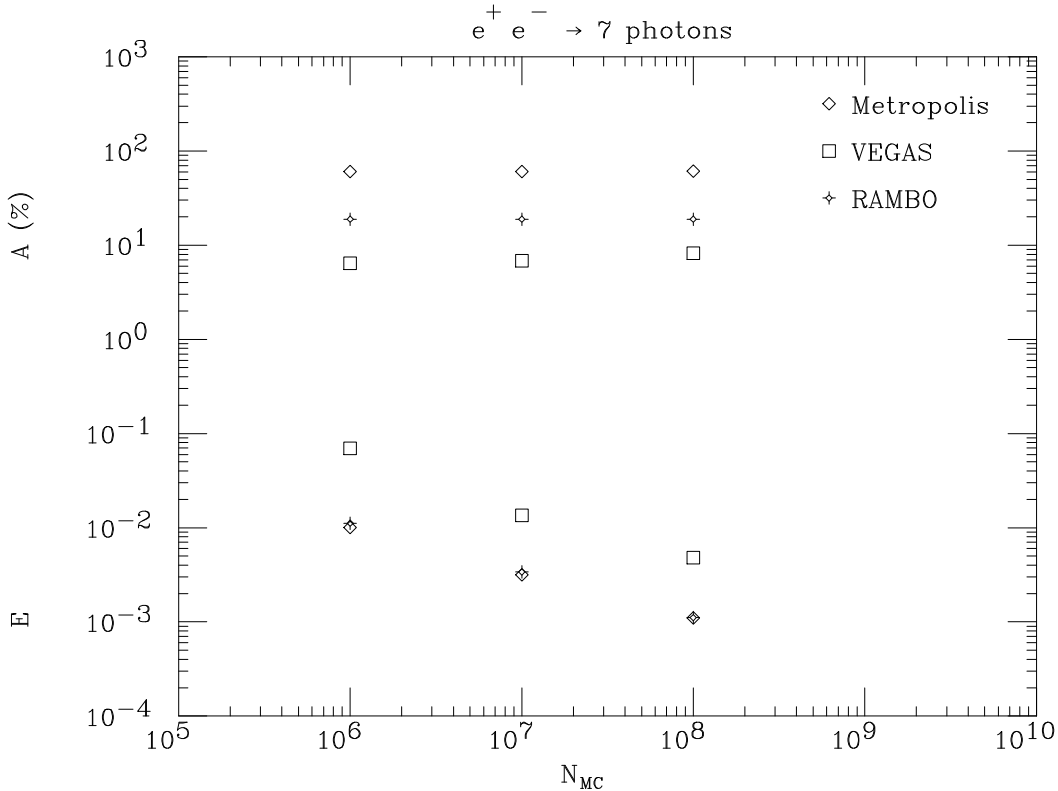}
\caption{Acceptance rates, $A$ (above), defined as the ratio of accepted over
generated MC points (times 100), and relative error, $E$ (below), defined as
the ratio of the standard deviation over the average value, 
for the Metropolis, {\tt VEGAS} and {\tt RAMBO} algorithms in evaluating
the total cross section for $e^+e^-\to 7~\gamma$, as a function
of $\NMC$. For {\tt VEGAS}, the setup  
${\rm{ITMX}}=4$ and ${\rm{NCALL}}=\NMC/{\rm{ITMX}}$ was used.
(Note the overlapping errors for Metropolis and {\tt RAMBO}.)}
\label{fig:ee7ph}
\end{center}
\end{figure}

Therefore, for high-dimensionality phase spaces, Metropolis, and {\tt RAMBO} 
as well, seem to be much more accurate than {\tt VEGAS}. However, one might
well wonder what is the actual number of points used in the evaluation of the
integral, as the accuracy of the latter strongly depends upon it. It turns
out that the acceptance rate of {\tt VEGAS} 
(recall Fig.~\ref{fig:acceptance}) is very poor compared to that
of Metropolis and {\tt RAMBO}, which is about 61(59)[56]\% and 19(10)[4]\% 
for $n=7(8)[9]$, respectively. (In particular, the large difference
between the acceptance rates of Metropolis and {\tt RAMBO} for the case $n=9$  
should explain the much smaller error for the former.)  
Thus, it is not surprising to see a bigger error in the former. Indeed,
if {\tt VEGAS} itself is run in non-adaptive mode (i.e., ${\rm{ITMX}}=1$)
its acceptance significantly improves (see again Fig.~\ref{fig:acceptance})
and the error consequently diminishes (typically halved), still being
larger than in Metropolis and {\tt RAMBO}, though. 

However, in order to show that the higher accuracy for Metropolis, as compared
to {\tt VEGAS}, for a large number of dimensions is not an artifact
due the specific value adopted for $\NMC$, 
we plot in Fig.~\ref{fig:ee7ph} both the acceptance and the
size of the relative error for the two algorithms
in calculating, e.g., $\sigma(e^+e^-\to 7~\gamma)$,  
with $\NMC=10^6, 10^7$ and $10^8$ (as usual, ${\rm{ITMX}}=4$ in {\tt VEGAS}).
Similarly, we proceed for {\tt RAMBO}. One can see that, 
no matter how many MC points
one can dispose of in {\tt VEGAS}, both the adaptability and accuracy
in Metropolis remain significantly better. The gain for {\tt VEGAS} is in the 
end only appreciable against {\tt RAMBO}: Metropolis still stands out
as the best choice for large $n$ values, whatever $\NMC$ is actually used
for the integration. This statement remains true for any choice 
${\rm{ITMX}}=1,2,4,5,10$ adopted in {\tt VEGAS}.

\subsection{Mass singularities, multi-channels and negative weights}
\label{eebbww}

The physics concerned with our discussion below can be found in
Refs.~\cite{split1,split2}. The process calculated is $e^+e^-\to
b\bar b W^+W^-$, with massive quarks (and gauge bosons, of course),
proceeding at tree-level through the 61 Feynman diagrams depicted
in Fig. 1 of
Refs.~\cite{split1,split2} (again, we assume $m_e=0$). These include several 
graph subsets (eight of these were isolated in Ref.
\cite{split2}), each having a peculiar (non-)resonant
structure, so that they 
 can be regarded as separate production modes 
of an actual multi-channel process. Furthermore, since
interference terms exist in the full ME among the various channels, 
some of the latter can give rise to 
negative contributions in the
integration procedure (the ${\cal T}$ weights of Metropolis discussed 
previously). 

For sake of illustration and comparison among the algorithms, 
rather than generating the total $e^+e^-\to b\bar b W^+W^-$ 
cross section in a unique run
using {\sl a-priori weights}  to choose
among the various channels (i.e., a l\`a \cite{kleiss}), 
we instead perform a separate integration over each of the latter, as
they present different challenges to the algorithms, see
Refs. \cite{split1,split2}.
In general, our approach can be viewed  as a preliminary by-hand optimisation
of the weights eventually
used in a full ME multi-channel run, rather than the automatic one 
discussed in Ref.~\cite{kleiss}.
Numerical values used here for the various parameters needed for the
calculation are as follows (the reader should not mind their obsolescence, as
they are used for illustration purposes): $\sqrt s=300$ GeV, $m_t=145$ GeV,
$\Gamma_t=0.78$ GeV, $m_b=5$ GeV, $M_H=120$ GeV, $\Gamma_H=6.9$ MeV,
$M_Z=91.1$ GeV, $\Gamma_Z=2.5$ GeV, $M_W=80$ GeV and $\Gamma_W=2.2$ GeV.
Note that we do not impose any cuts on the phase space (so that
all $\NMC$ generated points are actually used in the ME evaluations)
and we
neglect initial state radiation (ISR), thus identifying the partonic
energy with that of the collider.

\begin{table}[htb]
\begin{center}
\begin{tabular}{|c||c|c|c||c|}
\hline
\multicolumn{5}{|c|}
{\rule[0cm]{0cm}{0cm}
$\sigma (e^+e^-\to b\bar b W^+W^-)$ (fb) at $\sqrt s=300$ GeV}
\\ \hline\hline
channel & Metropolis & {\tt VEGAS} & {\tt RAMBO} & $\;$ \\ \hline
${\cal{T}}_1$          &  
$598.211\pm44.$        &          
$582.541\pm0.26$       &          
$517.794\pm56.$        &          
                                  \\ \hline  
${\cal{T}}_2$          &  
$2.94538\pm0.12$       &          
$2.78886\pm0.019$      &          
$2.66227\pm0.64$       &          
                                  \\ \hline  
${\cal{T}}_3$          &  
$2.85556\pm0.14$       &          
$2.79228\pm0.019$      &          
$3.82429\pm0.61$       &          
                                  \\ \hline  
${\cal{T}}_4$          &  
$1.96508\pm0.041$      &          
$1.85213\pm0.0013$     &          
$1.82016\pm 0.082$     &          
$\times10^{-2}$                   \\ \hline  
${\cal{T}}_5$          & 
$4.99161\pm0.079$      &          
$5.08556\pm0.0044$     &          
$4.78105\pm0.094$      &          
$\times10^{-1}$                   \\ \hline  
${\cal{T}}_6$          &  
$-1.21966\pm0.023$     &          
$-1.21143\pm0.0094$    &          
$-1.24077\pm0.035$     &          
$\times10^{-6}$                   \\ \hline  
${\cal{T}}_7$          &  
$4.80077\pm0.51$       &          
$4.69457\pm0.0043$     &          
$4.30172\pm1.1$        &          
$\times10^{-1}$                   \\ \hline  
${\cal{T}}_8$          &  
$1.34418\pm0.081$      &          
$1.20476\pm0.064$      &          
$1.14697\pm0.21$       &          
$\times10^{-1}$                   \\ \hline\hline
\multicolumn{5}{|c|} 
{\rule[0cm]{0cm}{0cm}
No cuts 
\qquad\qquad\qquad\qquad\qquad\qquad\qquad\qquad\qquad\qquad
No ISR}
\\ \hline
\end{tabular}
\caption{Contributions to the total cross section for $e^+e^-\to
b\bar b W^+W^-$ of the eight (non-)resonant channels of Ref.~\cite{split2}, as
obtained from the three algorithms documented in the 
text. For {\tt VEGAS}, the setup ${\rm{ITMX}}=5$ and ${\rm{NCALL}}=100000$
was adopted. }
\label{tab:Ti}
\end{center}
\end{table}

The algorithms chosen for this test are Metropolis, {\tt VEGAS} and 
{\tt RAMBO}. 
In running {\tt RAMBO}, we have not set up any special arrangement in
dealing with the complicate phase space structure of the various channels
\cite{split1,split2}:
we have let the algorithm generate 
four-momenta and weights and computed the estimate and 
error of the total cross section as described in the previous Subsection.
As for {\tt VEGAS}, we have adopted here
the same mapping of the integration variables
described in Ref.~\cite{split2}. Concerning the Metropolis implementation,
the expression in eq. (\ref{parti3}) of Subsect.~\ref{algdesc} was used 
for computing the cross section.
This method is applicable when the ME can be expected to have
negative values. Also, the massive phase space volume was
calculated beforehand and with insignificant errors.
Notice that the \NMC\ statistics used here should be taken
as representative of a value for which all three algorithms converge
to the correct integrals (see Ref.~\cite{split1,split2}). Indeed, if this
is augmented, errors diminish considerably in each case, though the relative
performances among Metropolis, {\tt VEGAS} and {\tt RAMBO} remain basically
unaffected.

The eight terms ${\cal{T}}_i$, $i=1, ... 8$, of eqs.~(9)--(16) in
Ref.~\cite{split2},  integrated
over the full four-particle phase space, are presented in Tab.~\ref{tab:Ti},
as obtained 
by using Metropolis, {\tt VEGAS} 
(${\rm{ITMX=5}}$ and ${\rm{NCALL=100000}}$)
and {\tt RAMBO}, with $\NMC=500000$. Here, it definitely is {\tt VEGAS}
to come out best, with second choice Metropolis and last {\tt RAMBO}.
The flaws of the latter should have been expected, as the algorithm is not
adaptive so that it suffers from the presence of peaks rising over the 
phase space. This is particular evident in the ${\cal{T}}_7$ channel,
which accounts for the very narrow $H\to b\bar b$ resonance \cite{split2}
(recall that the Higgs width is just a few MeV)\footnote{Further
notice that, being $M_H<2M_W$, the Higgs decay channel $H\to W^+W^-$
in ${\cal{T}}_4$ is not open.}. In fact, 
when the resonant particles involved have
a width of a few GeV
(i.e., $t$, $\bar t$, 
$Z$ and $W$), such as in ${\cal{T}}_i$ with $i=2, ... 6$, the
accuracy improves, unless two resonances have to be evaluated at once, those
from top quark pairs in ${\cal{T}}_1$. Here,
the error does become very significant (about 10\%). 

Metropolis behaves better than {\tt RAMBO}, as its error is always smaller.
It deals with single resonances rather satisfactorily for the quite
low statistics used (when $i=4,7$).
No particular problems arise in Metropolis in dealing with negative
weights (and rapidly changing interferences) either, i.e., ${\cal{T}}_i$ 
when $i=2,3,5,6,8$, as here the typical error
does not worsen in comparison to the cases in which the integrand function
is definite positive (i.e., $i=4,7$). However, Metropolis is no matching
to {\tt VEGAS}, particularly when multiple resonances are present, as
in ${\cal{T}}_1$. 

In the end, the careful mapping performed in {\tt VEGAS}
of all resonances and interferences has paid off. However, one should notice 
the minimal involvement of the Metropolis implementation in this
case, compared to the
{\tt VEGAS} one. 
in the Metropolis algorithm, there are no phase space Jacobian
factors to be accounted for. On the other hand, we have stressed
how they can efficiently be used in {\tt VEGAS}
to remove poles arising from the ME.
Indeed, we will show in the next Subsection that, 
if an effort  similar to that devoted to {\tt VEGAS} here is employed for 
Metropolis as well (in `teaching' to the
algorithm the singular structure of the integrand function), then
the performance of the latter can match that of the former.

\subsection{Variable energy,  
high multiplicity final states and cuts}
\label{hadron}

We dedicate this final Subsection of our numerical analysis to study
the process $gg\to b\bar b t\bar t$ $\to$ $b\bar b b\bar b W^+W^-$ $\to$
$b\bar b b\bar b {\mathrm {jj}} \ell^\pm \nu_\ell$ (where
j represents any light-quark jet and $\ell^\pm$/$\nu_\ell$ a lepton/neutrino),
 which was considered in 
Ref.~\cite{charged} as a QCD background to a possible 
charged Higgs discovery channel for the Large Hadron Collider (LHC).
Numerical parameters and other inputs used for the runs were declared there.

In this test, there are three specific technical problems associated 
with the calculation of the total cross section. Firstly, the fact that
the CM energy at partonic level is no longer a constant (contrary to the
two previous examples): being a hadron-hadron process,  
two more integrations (in additions to those over the phase space)
have to be performed, over the gluon momentum fractions, which evolve
according to the PDFs inside the proton.
Secondly, the very large number of final state particles, eight in total, which
imposes a 21-dimensional integration over the phase space\footnote{One degree
of freedom being absorbed into the flat azimuthal integration about
the incoming beam direction (see discussion
in Subsect.~\ref{photons}), and already accounting for the PDF convolution.}
(beside the presence of various mass singularities, both infrared and resonant
poles).
Thirdly, the introduction
 of severe reductions of the original integration region, 
as we have enforced in our simulations the same acceptance and selection cuts 
recommended in Ref.~\cite{charged}, the latter being implemented simply
through theta functions\footnote{In fact, it turns out impossible to map 
the entire phase space in terms of the kinematic quantities whose range is 
being cut, and not any more efficient to use only one or two 
of these as integration variables.}.

The algorithms used in this example were only two: Metropolis
and {\tt VEGAS}. The latter uses as usual a careful mapping around the
heavy particle resonances, through the variable 
\begin{equation}\label{atan}
\phi={\mathrm {arctan}}
\left( \frac{Q^2-M^2}{M\Gamma} \right)
\end{equation}
(where $M$ and $\Gamma$ represent the natural mass
and width of the unstable particle, $t,\bar t$ or $W^\pm$,
with virtuality $Q^2$),
whose derivative is proportional to the resonant propagator itself:
\begin{equation}\label{prop}
d\phi=\frac{M\Gamma}{(Q^2-M^2)^2+ M^2 \Gamma^2 }dQ^2.
\end{equation}
The setup of the former is as described in Subsect. 
\ref{varpar}. This is the most involved implementation of the Metropolis
algorithm which was tested. In this case, one needs an approximate 
ME with known cross section at a certain energy: this 
was constructed by simply using the two $W$- and the two $t$-resonances.
Therefore, the Metropolis approach adopted here
can be viewed as the equivalent of the {\tt VEGAS} mapping
enforced through eqs.~(\ref{atan})--(\ref{prop}).   The 
cross section associated to this auxiliary ME 
at one energy was calculated numerically beforehand. 
In the following, we assume the latter to be known with arbitrary small 
error. 
In this respect, we should also mention that in our actual ME we have ignored
interference effects between the two subsets of Feynman diagrams that
only differ in the exchange of the
four-momenta and spins between the two $b$-quarks (or, equivalently, the two
$b$-antiquarks) in the final state, because of their indistinguishibility 
in the detector,  
and a minus sign, because of the Fermi-Dirac statistics (in other terms, 
the Pauli principle). In fact, their effects on the total and differential
cross sections are negligible \cite{charged}. Besides, their integration
would pose further, unnecessary complications. 

\begin{table}[htb]
\begin{center}
\begin{tabular}{|c||c|c|}
\hline
\multicolumn{3}{|c|}
{\rule[0cm]{0cm}{0cm}
$\Delta\sigma/\sigma 
(gg\to X\to b\bar b b\bar b {\mathrm {jj}} \ell^\pm \nu_\ell)$ (\%)}
\\ \hline
$\sqrt{\hat s}$ (TeV) & {Metropolis} & {\tt VEGAS} \\ \hline\hline
\multicolumn{3}{|c|}
{\rule[0cm]{0cm}{0cm}
$X\to$ only top-antitop radiation, no PDFs}
\\ \hline
0.6         &  
$1.92$      & 
$0.48$        
            \\ \hline
1.0         &  
$4.25$      & 
$0.99$        
            \\ \hline
1.4         & 
$6.88$      & 
$1.52$        
            \\ \hline\hline
\multicolumn{3}{|c|}
{\rule[0cm]{0cm}{0cm}
$X\to$ full ME, no PDFs}
\\ \hline
0.6      &  
$0.77$   & 
$0.32$     
           \\ \hline
1.0      &  
$2.15$   & 
$0.61$     
         \\ \hline
1.4      & 
$2.96$   & 
$0.72$     
                     \\ \hline\hline
\multicolumn{3}{|c|}
{\rule[0cm]{0cm}{0cm}
$X\to$ full ME, PDFs}
\\ \hline
${\sqrt{\tau s}}$    & 
$3.58$      & 
$1.53$        
                     \\ \hline\hline
\multicolumn{3}{|c|}
{\rule[0cm]{0cm}{0cm}
Kinematics cuts from Ref.~\cite{charged}}
\\ \hline
\end{tabular}
\caption{The relative error on several cross sections associated to
the process
$gg\to X\to b\bar b b\bar b {\mathrm {jj}} \ell^\pm \nu_\ell$
at $\sqrt s=14$ TeV, 
as obtained from the two algorithms documented in the 
text, each using about $10^6$ MC points (all passing the default cuts
of Ref.~\cite{charged}).  
We have verified that actual cross sections (not shown here) 
are statistically consistent between the two algorithms.}
\label{tab:eight}
\end{center}
\end{table}

As we have already digressed to some length about the relative ability of
the two algorithms to adapt, we make our  
primary concern in this test that of comparing the size of the errors
associated to the integrals in each case. In order to render the
comparison consistent, regardless of the actual value of point generated,
\NMC, we always compute the integrals for a given statistics in both cases,
 e.g., $10^6$ (that is, the latter is the 
approximate number
of MC points that actually pass the cuts).
We proceed to obtaining the final result  by steps, in order to assess whether
 one  algorithm outperforms  the other in some specific task.
We start by isolating a gauge-invariant substructure of the original ME, 
only comprising
those diagrams (eight in total) in which the off-shell gluon, $g^*$, eventually
splitting into $b\bar b$ pairs, is emitted by either of the top (anti)quarks,
the latter finally decaying semileptonically. 
 Moreover, we fix
the partonic CM energy, i.e, $\sqrt{\hat{s}}=$
constant, thus removing for the time being the integrations over the gluon
PDFs. As the corresponding cross section has little meaning
physics-wise, we only plot the relative error as obtained from the
two algorithms, e.g., at $\sqrt{\hat s}=600,1000$ and 1400 GeV.
The upper part
of Tab.~\ref{tab:eight} shows the rates for the simpler 
subprocess just described, i.e.,
$gg\to t\bar t\to g^* t\bar t \to b\bar b t\bar t\to$ 
$b\bar b b\bar b W^+W^-$ $\to$
$b\bar b b\bar b {\mathrm {jj}} \ell^\pm \nu_\ell$. At all energies,
{\tt VEGAS} yields a smaller error than Metropolis, by about a factor
of four. 
Although Metropolis is still outperformed by {\tt VEGAS} in the size
of the relative error of the various integrations, one should notice
that the differences  between the two algorithms
have diminished substantially: compare to the rates in Tab.~\ref{tab:Ti}.
In both cases, the relative error increases with $\sqrt{\hat s}$.
In order to understand this effect, 
it should be recalled that, although the final state particles
are jets and leptons (thus with negligible rest mass as compared
to the $\sqrt{\hat{s}}$ values used), 
the two (anti)top resonances involved impose that the 
cross section would drastically fall to negligible levels if
$\sqrt{\hat{s}}\Ord 2(m_t+m_b)$.
Effectively, the volume of the phase space associated with the
integration performed at $\sqrt{\hat s}=600$ GeV is much smaller that
that spanned when $\sqrt{\hat s}$ is 1400 GeV, where the 
8-particle phase space 
can stretch much further away from the $b\bar b t\bar t$ threshold at 360
GeV. Therefore,
one would conclude that the tendency of both integrators to giving
smaller errors at lower CM energies is a consequence of the fact that 
 the integrand function fluctuates
more at larger
$\sqrt{\hat s}$.

As a second step of our test, we have introduced the full ME, 
see Ref.~\cite{charged}, in place
of the reduced one considered so far, where by `full' we intend
the one obtained by allowing for the attachments of the
off-shell $g^*\to b\bar b$ current to the gluon lines too, but 
with the process still proceeding
via the production of two top (anti)top quarks, and without the mentioned
interferences. This $2\to8$
ME consists of 36 basic Feynman diagrams.
Not to complicate things further, we again express the gluon
PDFs through $\delta$-functions centered around one and fix the
partonic energy $\sqrt{\hat{s}}$ at the usual 
three values: 600, 1000 and 1400 GeV. Results are presented in the
middle part of Tab.~\ref{tab:eight}. The relative performances
of the two algorithms are rather similar to the previous case, though the
overall error has diminished at each energy in both Metropolis and {\tt VEGAS}.
This effect is presumably the consequence of the fact that the
additional contributions to the ME fill regions of the phase
space previously empty, these smoothly interpolating into those
typical of the  kinematics of gluon emission from (anti)top quarks.

The final step is the integration over the full ME, including the convolution
with the PDFs, i.e., ${\hat{s}}=x_1x_2s=\tau s$, with $\sqrt s=14$ TeV.
In {\tt VEGAS}, the latter was done by adding the two 
integrations over the two gluon momentum fractions, i.e., $x_1$ and $x_2$,
 (so that ${\rm{NDIM}}=21$)
and the call to the PDF numerical package. In Metropolis, we have proceeded
as follows. One of the integrations over the momentum fractions was performed
beforehand. This was done by first 
changing the momentum fraction variables into the logarithm of the 
(squared) CM energy  and rapidity of the two gluons (as in
{\tt VEGAS}): 
$$
\frac{dx_1}{x_1}\frac{dx_2}{x_2}=
d\ln(x_1x_2)d\ln\sqrt{\frac{x_1}{x_2}}.
$$
The rapidity spectrum was then integrated at fixed CM energy, this yielding
a distribution in the latter variable only. This was in turn convoluted into 
the weight-function. The complete integration was finally 
done by combining the variable energy and
parallel integration methods, as described in Subsect. \ref{varpar}. The 
region $400~\mathrm{GeV}\Ord\sqrt{\hat s}\Ord 2$ TeV was chosen, 
large enough to gather the main contributions to the cross section. 
From the lower part of Tab.~\ref{tab:eight}, one can see that,
even in presence of the complete differential
structure of the cross section, 
{\tt VEGAS} performs better than Metropolis  in terms of accuracy,
but the relative size of the error has gone down, to a factor slightly
larger than two only. The absolute size is larger than in the 
previous test for both algorithms, a consequence of the additional 
dependence upon the gluon PDFs.

As intimated at the end of the previous Subsection,
if a similar sort of care is devoted to both algorithms 
in order to better account for the singular structure of the MEs,
one should conclude that, although {\tt VEGAS} is still better
in minimising the accuracy of the integration, 
Metropolis represents at least a viable alternative. This becomes 
particularly true if one finally considers that,
 even in this context, Metropolis displayed 
a better tendency than {\tt VEGAS} to adapt in high dimensions and
over a reduced phase space, as already seen elsewhere.
In other words, the same accuracy  can be achieved by 
the former with much less CPU-time effort than for the latter.


%% file: metroSumma.tex
Adaptive MC programs have become an indispensable tool in 
high energy particle physics, in order to calculate reliably
decay and scattering cross sections over multi-dimensional 
phase spaces. Indeed, as the energy reach of modern particle
accelerators grows larger, both the number of particles that can be
accommodated in the final state and that of the channels through which
they can be produced, increase rapidly. Under these circumstances, it is 
evident that simple generalisations of well known one-dimensional 
integration methods are no longer applicable, given the huge number of
points that would be needed in order to overcome the high complexity of the 
integrand functions (let alone the use of analytic methods, if one 
further considers the need of accounting for non-trivial reductions of the 
integration volume, because of unavoidable experimental cuts). 

Although several algorithms already exist on the market nowadays,
which  perform their task sufficiently well to have enabled stringent
tests between theory and experiments, 
it is clear that the availability of new implementations is a need
never exhausted: if only for cross-checking purposes. For this and other
reasons, we have developed several {\sl new} implementations of an {\sl 
old} MC technique: the so-called Metropolis method.
Although being another adaptive MC algorithm, it can boast at least
one radically different feature with respect to traditional approaches:
in the latter, the integration over the phase space takes place over a 
{\sl grid}; in the former, it evolves along a {\sl path}.
This discretisation of the integration volume has historically appealed
to the solution of problems in statistical mechanics and Lattice Gauge 
Theories. However, there is basically no reason why such a technique can not
be applied successfully also to the four-momentum phase space of Quantum Field
Theories. 

As a matter of fact, we have here demonstrated the high potential of the 
Metropolis method in dealing with realistic problems arising in modern
particle physics phenomenology. It not only fulfills the basic criteria of
accuracy and efficiency required to any algorithm, no matter the number of 
dimensions involved in the integration,  but it has also been shown to 
outperform in some cases 
other, already widely diffused MC programs. Moreover, 
being its speed matter of no concern at
all, the Metropolis algorithm could well be used in (parton level) MC 
event generators too.

There can certainly be drawbacks in the application of the method. The 
most important being the need at times of optimising the implementation of the 
Metropolis algorithm to solve a specific problem, which can require 
prior knowledge of the behaviour of the integrand function.
But even then, the reader should acknowledge that 
it is becoming more and more rare in numerical simulations
that one can afford to rely solely on the ability of whatever algorithm in
adapting itself to such subtle effects as interferences, finite particle
widths, irreducible backgrounds, etc., as some of our examples 
should clearly have demonstrated. 

We have encoded the various implementations of the Metropolis
algorithm discussed in this paper in a C++ program, that we make available
to the public upon request. This code makes extensive use of the CLHEP 
classes \cite{CLHEP} for handling Lorentz four-momenta.

Before closing, we will describe two possible 
improvements that could be considered in the future in order to
further increase the efficiency of the Metropolis algorithm in QFT. Both these
methods are widely used for conventional statistical physics problems, and
shown to be necessary tools in many cases. 

The first method relies on the introduction 
of a fictitious reciprocal temperature, $\beta$. This
method has seen two different implementations called `simulated tempering'
\cite{dyntemp} and `parallel tempering' 
\cite{partemp}\footnote{The
difference between the two lies essentially in the way the 
configuration space is enlarged by means of 
$\beta$.}. In our case, the phase space would be enlarged with
$\beta\in[0,1]$ as a new parameter, inducing a modified
 weight-function,
$$
w(x,\beta)=w^\beta(x).
$$
Clearly, the integration is simpler with $\beta$ close to zero, while we want
it instead to be performed at $\beta=1$. This way of enlarging the phase space
allows for round tours into the low-$\beta$ region where the configuration can
be updated more freely, and thus the correlation length is shorter. This is
very much in the same manner as when the total energy, $E_{\mathrm{tot}}$, was 
used as a
free parameter (see Subsect. \ref{ene} of this article). 
The advantage of instead
using $\beta$ is that the efficiency of the method would not be as much
dependent  on the energy variations of the cross section. In any case, this 
method should, and could rather easily, be tested also in the case
of integrations over a four-momentum phase space.

The other technique
 that we want to mention is known as the `Hybrid Monte Carlo'
method \cite{hybris}. In this case, derivatives of the weight function are
used to move large distances in configuration space, but still avoiding the 
un-favoured regions where the weight function is small. In many cases, this
method has increased the efficiency of the Metropolis algorithm drastically
and could very well be unavoidable if much more numerically demanding problems
are to be treated in the QFTs of the future. Unfortunately, derivatives of more
complicated matrix elements are not available today. It is also at present not 
clear to the authors how this hybrid MC should be implemented to 
perform integrations over a four-momentum phase space.

A final, more general remark, is the following. Throughout
 this article, we have described several implementations of the Metropolis
algorithm, all using a stepping procedure designed in such a way that the
final state particles are kept on mass-shell. An alternative approach would be
to map their four-momenta onto a set of independent variables and perform the
random walk in the new phase space. This way, one could benefit
from the use of mappings which can cancel the ME singularities (e.g., see
eqs. (\ref{atan}) and (\ref{prop}) in Sect. \ref{hadron}). A possible
implementation of this last strategy is now under consideration.
 

%% file: metroApp.tex
\section*{Appendix: a proof for restricted phase spaces}
\label{acond}

In this Appendix, we will show that, in case the integration is to be done in
a subspace of the original phase space, the $PS_{\rm{red}}$ of 
Subsect.~\ref{algdesc}\footnote{Which we assume 
here to be an {\sl open} subspace.},
 the Metropolis condition (a) is fulfilled, if for any two
points inside  $PS_{\rm{red}}$, there exists a continuous curve $\Gamma$, 
which connects the two points and  is contained in $PS_{\rm{red}}$.

For every point $x$ along $\Gamma$, there is an open disc $D(x,r)$, with 
radius\footnote{With the metric  of the
Euclidian space of the three-momenta.} $r$, which is wholly contained in
$PS_{\rm{red}}$. First, we show that we can choose a $\delta$, 
$0\!<\!\delta\!<\!r$, so that all points on the surface $\{y;|x-y|=\delta\}$
can be reached from $x$ in $n-1$ steps which are all inside $D$.

Suppose, as before, that $x=\{p_i\}$ and $y=\{p'_i\}$ and let 
$\Delta_i=p'_i-p_i$. We can now move from $x$ to $y$ in $n-1$ steps, by 
transferring the $\Delta_i$ one by one to $p_n$ for $i=1$ to $i=n-1$. This
will  take us through $n-1$ points: 
$x=x_0\rightarrow x_1\rightarrow \dots \rightarrow x_{n-1}=y$. 
Let $d$ be the largest of $|x-x_i|$ in this path.
Choose an $a$ such that $0\!<\!a\!<\!r/d$ and get a new
$\delta'=a\delta$. Then the point $y'\equiv x+a(y-x)$ lies on the corresponding
surface and can be reached by a path inside the disc $D$. If $\tilde y$ is
the point on $\{y;|x-y|=\delta\}$ which gives the largest $d=\tilde{d}$ and
with $0\!<\!\tilde{a}\!<\!r/\tilde{d}$, we
have that all points in $S=\{y;|x-y|=\tilde{a}\delta\}$ can be reached in a
finite number of steps, which are all contained in $D$.

A point where $\Gamma$ crosses $S$ has a finite distance ($\tilde{a}\delta$)
to $x$ and it can then be reached with a finite number of
steps, all inside $PS_{\rm{red}}$. In this way, starting from any point 
in $PS_{\rm{red}}$,
any other point can be reached with a finite number of steps inside
$PS_{\rm{red}}$, by taking finite steps along the corresponding $\Gamma$.
This concludes our proof. 


%% file: error.tex
\begin{titlepage}

  \begin{flushright}
    {DESY 99-133E}\\
    RAL-TR-1999-061E\\
    TSL/ISV-99-0216E\\
    May 2000
  \end{flushright}
  \begin{center}

    \vskip 5mm
    {\LARGE\bf The Metropolis algorithm for on-shell\\[0.35cm]
      four-momentum phase space: Erratum}
    \vskip 5mm

    {\large Hamid Kharraziha}\\[0.25cm]
    {\it II. Institut f\"ur Theoretische Physik}\\
    {\it Universit\"at Hamburg}\\
    {\it D-22761 Hamburg, Germany}\\
    hamid.kharraziha@desy.de\\
    \vskip 10mm

    {\large Stefano Moretti}\\[0.25cm]
    {\it Rutherford Appleton Laboratory}\\
    {\it Chilton, Didcot, Oxon OX11 0QX, UK}\\
    moretti@v2.rl.ac.uk\\[3mm]
    and\\[3mm]
    {\it Department of Radiation Sciences}\\
    {\it Uppsala University, P.O. Box 535}\\
    {\it 75121 Uppsala, Sweden}\\
    moretti@tsl.uu.se

  \end{center}
  \vskip 10mm
  \begin{abstract}
    \noindent
A proof presented in the original paper was incorrect.
We outline here an alternative procedure.
  \end{abstract}

\end{titlepage}

\section*{The new proof}

One of the conditions required to the stepping procedure implemented in the
Metropolis algorithm, as described in  
Ref.~\cite{metro} (point (a) in Sect.~2.1), was that {\sl any point 
$\{p_i'\}$ in the $n$-particle phase space can be reached from any other
point $\{p_i\}$ in a finite number of steps}. The proof in the main article
relies on the false statement that, for any two on-mass-shell
particles with four-momenta
$p_1$ and $p_2$, the
whole of the surface of conserved $p_1+p_2$ can be reached by the 
identical particles by rotating them in
their centre-of-mass
(CM) frame. This is however not true, because it would in general be 
possible to transfer a four-momentum which conserves $p_1+p_2$ but puts 
one or both particles out of the mass-shell. (The rotation procedure itself, 
in contrast, does preserve the on-mass-shell condition). Below we
present a correct proof of the above statement.

It is clear that the whole of the two particle phase space can be reached in 
a single step, so let us assume that the number of particles 
($n$) is at least three. We will later on show that it is possible to move any 
particle (e.g., the $n$:th), from its original four-momentum $p_n$ to the final one
$p_n'$ in a finite number of steps, by exchanging some momentum with the other 
particles. This procedure will change the other four-momenta to new values 
$\{p_i''\}$. The rest of the problem is then to move these to their final
values $\{p_i'\}$. For $(n-1)=2$, this can be accomplished in one step. 
For $n>3$, we can move the $(n-1)$:th four-momentum to its final value $p_{n-1}'$, 
and so on.

Let us first discuss what four-momenta are possible for a given particle, in
the case of unrestricted phase space. Consider again the
$n$:th particle, and let $s$ be the total invariant energy squared, 
$s_{n-1}$ be the energy squared of the system of all the other particles and 
let $m_n$ 
be the mass of the n:th particle. In the CM frame, the magnitude of the 
three-momentum of the $n$:th particle is given 
by\footnote{Hereafter, 
we use $P$ to indicate the magnitude of a momentum in the CM frame
  and $p$ to denote a four-vector in any frame.}
$$
P_n=\frac{1}{2\sqrt{s}}\sqrt{\lambda\left(s_{n-1},m_n^2,s\right)},
$$
with
$$
\lambda(a,b,c)=a^2+b^2+c^2-2ab-2bc-2ca,
$$
and it can have any direction. Depending on the value of
$s_{n-1}$, $P_n$ can have any value between 0 
(for $s_{n-1}=(\sqrt{s}-m_n)^2$) and its maximum, $P_n^{\rm{max}}$, which 
occurs at $s_{n-1}=(\sum_{i=1}^{n-1}m_i)^2$ (the minimal value of $s_{n-1}$).

The four-momentum $p_n$ is, from its mass-shell condition, restricted to a
three-dimensional sub-space. We will now show that, in case $P_n$ is 
smaller that its maximal value, there exists a 
three-dimensional on-mass-shell neighbourhood of $p_n$ in which any point
can be reached in two steps, by exchanging momentum with two other particles 
(denoted below by $p_1$ and $p_2$). 

Let $k_1$ denote the exchanged four-momentum from $p_1$ to $p_n$. Our 
condition for the exchanged momentum is that it keeps both of the particles 
on their respective mass-shell. Consequently, with $p_n'=p_n+k_1$ we have
$p_n^2=p_n^{'2}=p_n^2+k_1^2+2k_1p_n \Leftrightarrow k_1^2=-2k_1p_n$.
Similarly, we have the condition $k_1^2=2k_1p_1$, in order to keep $p_1$ on the
mass-shell. These two conditions restrict $k_1$ to lie on a two-dimensional
surface (corresponding to the two rotational degrees of freedom in the CM 
frame of $p_n$ and $p_1$.) We will now show that two
particles $p_1$ and $p_2$ can be chosen so that the corresponding
infinitesimal four-momentum transfers $\epsilon k_1$ 
and $\epsilon k_2$ ($\epsilon\rightarrow 0$) to $p_n$
are such that $k=\epsilon k_1+\epsilon k_2$ has three degrees of freedom. 
This implies that $p_n+k$ (for a finite $k$) covers an open 
on-mass-shell neighbourhood of $p_n$. In the limit $\epsilon\rightarrow 0$, 
the on-shell conditions impose
\begin{eqnarray}
p_nk_1&=&0, \nonumber \\
p_nk_2&=&0, \nonumber \\
p_1k_1&=&0, \nonumber \\
p_2k_2&=&0. \nonumber
\end{eqnarray}
Under such conditions, the four-vectors $k_i$ are restricted to lie in 
two-dimensional planes. The vector $k$ will then have more than two degrees of 
freedom, unless the planes for $k_1$ and $k_2$ coincide. This would in turn
imply that the four-momenta
$p_1$ and $p_2$ are linearly dependent, that is, one can write $p_1=ap_2$ for
a number $a$. Consequently, it is enough to show that there exist two 
other particles such that their four-momenta are not proportional. This
condition is equivalent to say that $P_n$, as was assumed before, is smaller than 
its maximal value. 

To demonstrate this statement, we will show that $s_{n-1}$ reaches its minimum
if and only if all $n-1$ four-momenta are proportional to each other, that
is, there are numbers $a_{ij}$ so that one can write $p_i=a_{ij}p_j$, for
any $i,j\le n-1$. This is well known in case all particles are mass-less
($s_{n-1}=0$). In case any of them has a mass, one can boost the entire
system to the rest frame of such a particle. The minimal value of $s_{n-1}$ is 
then reached if
all particles have zero three-momentum in the new frame. 

This completes the proof that, in case
$P_n<P_n^{\rm{max}}$, there is a three-dimensional neighbourhood of $p_n$ which 
can be reached in two steps. From this, we draw the conclusion that the 
final point $p_n'$ can be reached in a finite number of steps, in case both 
$P_n$ and $P_n'$ are smaller than the maximum. If this was not the case, then 
there would exist a four-momentum $p''$, with $P''<P_n^{\rm{max}}$, which $p_n$
 can 
come arbitrarily close to, but never reach. But this contradicts the fact 
that a neighbourhood of $p''$ is reachable from $p''$ (and vice versa).

The final step is to show that it is always
possible (in case $n>2$) to move from 
a point with $P_n=P^{\rm{max}}_n$ to another point with $P_n'<P^{\rm{max}}_n$.
 This
follows from the condition $p_i=a_{ij}p_j$, that leaves one degree of freedom
for the four-momentum of one of the particles, for fixed values of the 
others. Since the four-momentum transfer $k$
has two degrees of freedom, one can, in one step, violate the proportionality 
condition above, with the consequence that $P_n'<P^{\rm{max}}_n$.